# Atomic-referenced Hz-linewidth lasers via fiber interferometric stabilization

Changmin Ahn[1†], Hansol Jeong[2†], Seoyeon Yang[2], Junyong Choi[1], Igju Jeon[1], Hanseb Moon[2*], and Jungwon Kim[1*]

[1]*Korea Advanced Institute of Science and Technology (KAIST), Daejeon 34141, Korea*

[2]*Pusan National University, Busan 46241, Korea*

†*These authors contributed equally*

* *e-mails: jungwon.kim@kaist.ac.kr; hsmoon@pusan.ac.kr*

**Narrow-linewidth lasers with absolute frequency anchoring are essential for precision metrology, coherent sensing, and emerging quantum technologies beyond laboratory environments. Optical cavities and interferometers provide exceptional short-term spectral purity but lack intrinsic absolute frequency references. Atomic transitions, in contrast, provide stable frequency anchors but offer limited discrimination sensitivity. Recent hybrid approaches have demonstrated the combination of compact optical resonators with atomic references, yet achieving the Hz-level regime remains challenging. Here, we present a hybrid architecture that enables simultaneous realization of Hz-level linewidth and atomic-referenced frequency stability. An external-cavity diode laser is first stabilized to a fiber interferometer to achieve Hz-level spectral purity, while the interferometer is subsequently anchored to an $^{87}$Rb $D_2$ transition via modulation transfer**

**spectroscopy to suppress long-term drift and define the laser frequency relative to the atomic transition. This dual-stabilization scheme realizes a compact atomic-referenced laser with a 3.4-Hz linewidth (1-rad integrated-phase method), a minimum fractional frequency stability of 3.4×10$^{-14}$ at 0.56 s, and 9×10$^{-13}$ at 100 s. This architecture establishes a practical and scalable route toward compact and field-deployable atomic-referenced narrow-linewidth lasers for precision metrology and quantum technologies.**

## 1. Introduction

Optical atomic references[1] have enabled unprecedented levels of frequency precision and stability, reaching fractional instabilities at the 10$^{-18}$ level in state-of-the-art ion[2,3] and optical lattice clocks[4,5]. These systems have become indispensable for applications ranging from the redefinition of the SI second[6-8] to tests of fundamental physics[9-11] and relativistic geodesy[12,13]. However, their large footprint, complex optical subsystems and stringent environmental requirements have largely confined their operation to specialized laboratory environments, limiting their deployment in practical and field-based settings.

To extend precision frequency references beyond the laboratory, compact atomic platforms have been actively developed[14-18]. While these systems provide robust frequency anchoring based on atomic transitions, their broad spectroscopic linewidths fundamentally limit frequency discrimination sensitivity. As a result, directly stabilizing a laser to such references does not provide sufficiently low short-term noise or Hz-level linewidth, which are essential for coherent applications. In many emerging applications, including quantum communications[19-21], quantum sensing[22-24], precision navigation[25,26] and distributed timing networks[27,28], both long-term stability and short-term phase coherence are simultaneously required.

In parallel, compact optical frequency references based on cavities[29-32] or fiber interferometers[33-36] have emerged as promising approaches for ultra-narrow linewidth generation. Optical cavities can provide excellent frequency stability, while fiber-based interferometers offer alignment-free operation and scalability through long optical delays. These platforms enable high frequency discrimination sensitivity and have demonstrated frequency stability approaching the thermal noise limit at short averaging times.

Despite these advances, a fundamental limitation remains. Interferometric and cavity-based systems lack an absolute frequency reference and are therefore susceptible to long-term drift, while atomic references provide stable frequency anchors but insufficient sensitivity for ultranarrow linewidth generation. To address this trade-off, hybrid approaches that combine compact optical references with atomic transitions have been reported[37,38]. However, existing implementations remain limited in reaching the Hz-level regime. In particular, reduced frequency discrimination sensitivity and thermorefractive noise hinder ultra-narrow linewidth generation, while their relatively large free spectral range (FSR) imposes constraints on resonance alignment to atomic transitions.

Here, we address this challenge by combining a fiber interferometer with atomic frequency anchoring in a compact platform. A kHz-linewidth external-cavity diode laser is first stabilized to a compact 1-km homodyne fiber interferometer, achieving Hz-level linewidth and low short-term noise. The interferometer is subsequently referenced to the $^{87}$Rb $D_2$ transition via modulation transfer spectroscopy[39], compensating for long-term drift by controlling the fiber delay length. In particular, fiber interferometers provide ultrahigh frequency discrimination sensitivity, allowing near-thermal-noise-limited stability at short averaging times without vacuum enclosures or complex free-space optics[34]. In addition, their small FSR provides practical advantages for matching the interferometric reference to the atomic transitions. As a result, we realize a 3.4-Hz linewidth atomic-referenced laser with a minimum fractional frequency stability of $3.4\times10^{-14}$ at 0.56 s and $9\times10^{-13}$ at 100 s, establishing a practical and scalable route toward compact atomic-referenced narrow-linewidth lasers for precision metrology and quantum technologies.

## 2. Results

### 2.1. Concept of atomic-referenced narrow linewidth laser

Figure 1**a** illustrates the conceptual difference between a conventional atomic-referenced laser and the proposed scheme. In conventional approaches, the laser frequency is directly locked to an atomic transition. While the intrinsic stability of atomic energy levels provides a well-defined absolute frequency reference, the relatively broad spectroscopic linewidth of compact atomic systems fundamentally limits the frequency discrimination sensitivity. As a result, direct locking to such references cannot achieve ultra-narrow linewidth or low short-term frequency noise.

To overcome this limitation, we decouple the roles of spectral purification and frequency referencing. As shown in Fig. 1**a**, a fiber interferometer is first introduced as a high-resolution frequency discriminator to suppress the laser linewidth, while the atomic transition is used to define the absolute frequency reference of the interferometer itself.

Figure 1**b** presents the operating principle of the proposed dual-stabilization scheme. The free-running laser is first stabilized to a 1-km fiber-delay Michelson interferometer by locking to the quadrature point of its transfer function. Owing to the long fiber delay, the interferometer provides high frequency discrimination sensitivity, enabling Hz-level

linewidth reduction and low short-term frequency noise.

However, the optical path length of the fiber interferometer is sensitive to environmental perturbations, such as thermal fluctuations and acoustic vibrations, which shift the locking point over time and degrade long-term frequency stability. To suppress this drift, the interferometer is referenced to an atomic transition by controlling its effective delay using a PZT fiber stretcher. In this configuration, the atomic transition does not directly stabilize the laser, but instead defines the absolute frequency anchor of the interferometric reference.

In contrast to free-space cavities, this approach enables alignment-insensitive operation and straightforward length actuation using fiber-integrated components. Furthermore, the relatively small FSR of the long-fiber interferometer significantly relaxes the wavelength tuning requirements for matching the interferometric fringe to the target atomic transition.

This hierarchical approach establishes a clear separation of roles: the interferometer determines the short-term spectral purity, while the atomic transition provides long-term frequency anchoring. As a result, the system simultaneously achieves ultra-narrow linewidth and atomic-referenced frequency stability, enabling compact atomic-referenced narrow-linewidth lasers.

### 2.2. Characteristics of the $^{87}$Rb $D_2$-line MTS reference

To establish an atomic frequency reference, a 1560.4-nm seed laser was frequency-doubled and stabilized to the $5S_{1/2}(F=2)\rightarrow 5P_{3/2}(F'=3)$ transition of $^{87}$Rb. Instead of saturated absorption spectroscopy (SAS), modulation transfer spectroscopy (MTS) was employed to generate a background-free dispersive error signal with a well-defined zero crossing as shown in Fig. 2**a** (see Methods for the detailed information). Figure 2**c** compares the measured SAS and MTS signals, highlighting the characteristic dispersive 'S-shaped' profile of the MTS signal. By locking the laser frequency to the zero-crossing of the MTS signal, an atomic-referenced laser was generated.

To evaluate the performance of the $^{87}$Rb $D_2$ line-referenced laser, the optical beat-note measurements were performed using two independent atomic-referenced lasers[40] (see Fig. 2**b** and Methods). The measured frequency noise power spectral densities (PSDs) of the free-running and atomic-referenced lasers are shown in Fig. 2**d**. After stabilization to the $^{87}$Rb $D_2$ transition, the frequency noise is significantly suppressed at low offset frequencies below ~300 Hz. For example, the frequency noise PSD at 10 Hz offset decreases from $2.9\times10^6$ to $4.7\times10^3$ $Hz^2/Hz$, corresponding to ~28 dB improvement. At higher offset frequency above ~300 Hz, the frequency noise approaches that of the free-running case. As a result, the linewidth estimated using the 1-rad integration method[41] is 1.6 kHz for the atomic-referenced laser, comparable to 1.3 kHz for the free-running case.

The long-term time trace of the beating frequency and Allan deviation are shown in Fig. 2**e** and **f**, respectively. In the free-running case, the root-mean-square (rms) beat-frequency drift is 1.2 MHz, whereas it is reduced to 324 Hz after stabilization, corresponding to an improvement by a factor of ~4000. Consistently, the frequency stability improved from $3\times10^{-9}$ to $9.6\times10^{-13}$ at 100 s averaging time, indicating that atomic referencing effectively suppresses long-term frequency drift while providing only limited improvement in short-term linewidth.

### 2.3. Atomic-referenced narrow-linewidth laser

Figure 3**a** illustrates the experimental implementation of the atomic-referenced narrow-linewidth laser. The 1560.4-nm laser is divided into two branches: one for interferometric stabilization and the other for atomic referencing. In the first branch, a 1-km fiber-delay Michelson interferometer, with an FSR of ~100 kHz and a corresponding Q-factor of ~$4\times10^9$, is used as a high-resolution frequency discriminator to suppress short-term frequency noise and reduce the laser linewidth (see Methods for details).

In the second branch, the frequency of the fiber-stabilized laser is compared with the $^{87}Rb$ $D_2$ transition using an MTS system. The resulting error signal is applied to a PZT fiber stretcher to control the interferometer delay, thereby referencing the interferometer to the atomic transition and compensating slow environmental drift. To prevent the relatively high short-term noise of the MTS system from degrading the linewidth, a low-pass filter with a 5-Hz cutoff frequency is introduced in the atomic-referencing loop. The performance of the resulting laser is evaluated using optical beat-note measurements between two independent systems (see Fig. 3**b** and Methods).

Figure 4**a** shows the frequency-noise PSDs under various conditions. At a 10-Hz offset frequency, the proposed laser exhibits a PSD of 1.9 $Hz^2/Hz$, corresponding to a ~34 dB reduction relative to the atomic-referenced laser. Importantly, the noise closely follows that of the fiber-stabilized laser over a wide frequency range, indicating that the ultra-narrow linewidth performance provided by the interferometer is preserved after atomic referencing. At offset frequencies above 20 kHz, the PSD is limited by resonance peaks associated with the interferometric locking loop, consistent with the in-loop response. The linewidth, estimated using the 1-rad integrated-phase method, is reduced to 3.4 Hz, compared to 1.3 kHz and 1.6 kHz for the free-running and atomic-referenced lasers, respectively, corresponding to more than two orders of magnitude reduction.

The long-term performance is shown in Fig. 4**b**. When stabilized only by the fiber interferometer, the rms frequency drift reaches 134 kHz due to environmental perturbations. In contrast, referencing the interferometer to the atomic transition reduces this drift to 474 Hz, an improvement by a factor of ~280. This value is comparable to that of the atomic-referenced laser (313 Hz), indicating that the long-term stability is effectively inherited from the atomic reference.

The corresponding Allan deviation is shown in Fig. 4**c**. At short averaging times, the stability follows that of the fiber-stabilized laser, while at longer times it approaches that of the atomic-referenced laser. For example, at 1 ms, the proposed laser exhibits a stability of $2\times10^{-13}$, ~60 times lower than that of the atomic-referenced laser ($1.2\times10^{-11}$). The stability reaches a minimum of $3.4\times10^{-14}$ at 0.56 s. In contrast, the fiber-stabilized laser diverges beyond ~0.01 s and follows an approximate $\tau^1$ dependence due to environmental drift. The proposed scheme suppresses this long-term drift and maintains a stability of $9\times10^{-13}$ at 100 s, approaching the long-term performance of the atomic-referenced laser. These results confirm that referencing the interferometer to the atomic transition enables simultaneous realization of ultra-narrow linewidth and long-term frequency stability.

Figure 4**d** places the performance of the proposed system in the context of previously reported compact atomic-referenced and cavity-based lasers[37,38]. The proposed laser clearly operates in a distinct performance regime, exhibiting orders-of-magnitude lower frequency noise and enhanced stability compared to prior works, while simultaneously achieving Hz-level linewidth and atomic-referenced stability.

## 3. Conclusion

The proposed architecture is not restricted to the $^{87}$Rb $D_2$ transition demonstrated here.

Because the interferometric reference is defined independently of the atomic system, the same concept can be extended to a wide range of compact atomic references[42], including two-photon Rb[16,43] and iodine-based systems[17,44]. This highlights that the approach is not tied to a specific transition, but instead provides a flexible framework for combining interferometric spectral purification with atomic frequency anchoring.

Beyond single-wavelength operation, the stabilized fiber interferometer can serve as a transfer reference[45,46] for additional lasers within the bandwidth of the fiber platform. Once anchored to an atomic transition, the interferometer provides a common frequency reference that can be distributed to multiple optical channels. Combined with nonlinear frequency-conversion techniques such as second-harmonic generation (SHG), this approach offers a practical route for extending atomic-referenced stability across a broad spectral range and enabling coherent multi-wavelength optical sources referenced to a single atomic transition.

In summary, we demonstrate a compact platform that bridges interferometric precision with atomic frequency anchoring to achieve Hz-level linewidth together with long-term stability. The system exhibits a linewidth of 3.4 Hz, a minimum fractional frequency stability of $3.4\times10^{-14}$, and $9\times10^{-13}$ at 100 s. This approach effectively separates spectral purification from frequency referencing, enabling independent optimization of short- and long-term stability within a single compact platform.

More broadly, this approach provides a pathway toward compact, coherent, and atomic-referenced optical sources. Such systems are expected to support applications including portable optical clocks[14,15], next-generation communication systems[19-21], quantum sensing[22-24], and quantum information processing[47-49], where both ultrahigh coherence and long-term frequency stability are required.

## 4. Methods

**$^{87}$Rb $D_2$-line modulation transfer spectroscopy (MTS):** A seed laser (ULN15TK, Thorlabs) operating at ~1560.4 nm was frequency-doubled to 780.2 nm via SHG to match the $^{87}$Rb $D_2$ transition. The second-harmonic light was delivered to the MTS setup and split into counter-propagating pump and probe beams. The pump beam was phase-modulated at ~5 MHz using an electro-optic modulator (EOM) driven by a function generator (FG 1 in Fig. 2**a**; EDU33212A, Keysight Technologies). Inside the Rb vapor cell, this modulation was transferred to the probe beam through a nonlinear four-wave mixing process. The pump-to-probe power ratio was adjusted using a half-wave plate and a polarizing beam splitter. The transmitted probe beam was detected by a photodetector, and a bias-tee was used to separate the AC component from the DC background. The AC signal was amplified and mixed with the local oscillator from FG 1 (in Fig. 2**a**), followed by low-pass filtering. The phase of the local oscillator was adjusted to maximize the MTS error signal. The resulting error signal was applied either to the laser frequency modulation port or to the PZT stretcher, enabling direct atomic referencing of the laser or stabilization of the fiber interferometer to the atomic transition, respectively.

**Fiber homodyne interferometer for interferometric stabilization:** The laser frequency was stabilized using an imbalanced 1-km fiber homodyne Michelson interferometer, corresponding to an effective optical delay of 2 km (with ~100 kHz FSR). Faraday mirrors were placed at the ends of both arms to suppress polarization-induced signal fading. A PZT fiber stretcher was inserted in one arm to enable stabilization of the interferometer length to the atomic reference. The fiber was wound onto a custom-designed spool to reduce sensitivity to mechanical vibration[50], and the entire interferometer was enclosed in a polycarbonate housing to mitigate acoustic and thermal perturbations. The interference signal was detected using a balanced photodetector (BPD) to suppress common-mode intensity noise. The resulting error signal was fed back to the laser current modulation port, stabilizing the laser frequency at the quadrature point of the interferometer transfer function.

**Beat-note measurement for performance evaluation:** To characterize frequency noise and long-term stability, beat-note measurements were performed using two independent and nominally identical laser systems. The optical outputs were combined using a 50:50 fiber coupler and detected by a photodetector. To avoid near-zero beat frequencies, one laser was frequency-shifted by 50 MHz using an acousto-optic frequency shifter (AOFS). The beat signal was divided by 32 to bring the carrier frequency within the measurement range. Phase noise was measured using a phase noise analyzer (5125A, Symmetricom) and converted to frequency-noise PSD. Time-domain measurements were recorded using a frequency counter (53200A, Keysight Technologies), from which the Allan deviation was calculated. Assuming identical and uncorrelated noise from the two systems, the measured frequency-noise PSD was divided by two, while the time trace and Allan deviation were divided by $2^{0.5}$.

**Funding**

National Research Foundation (NRF) of Korea (RS-2024-00334727, RS-2024-00436737, RS-2021-NR060086), Institute for Information and Communications Technology Promotion (IITP) of Korea (RS-2025-25464839, RS-2024-00396999, RS-2022-II221029,

IITP-2026-2020-0-01606)

**Data availability statement**

The data that support the findings of this study are available from the corresponding authors upon request.

**Competing interests**

J.K. and I.J. are inventors on patent applications related to this work filed by KAIST (Korean patent application 10-2024-0123922 filed on September 11, 2024, Korean patent application 10-2025-0125040 filed on September 3, 2025, and US patent application 19/317596 filed on September 3, 2025). The authors declare no other competing interests.

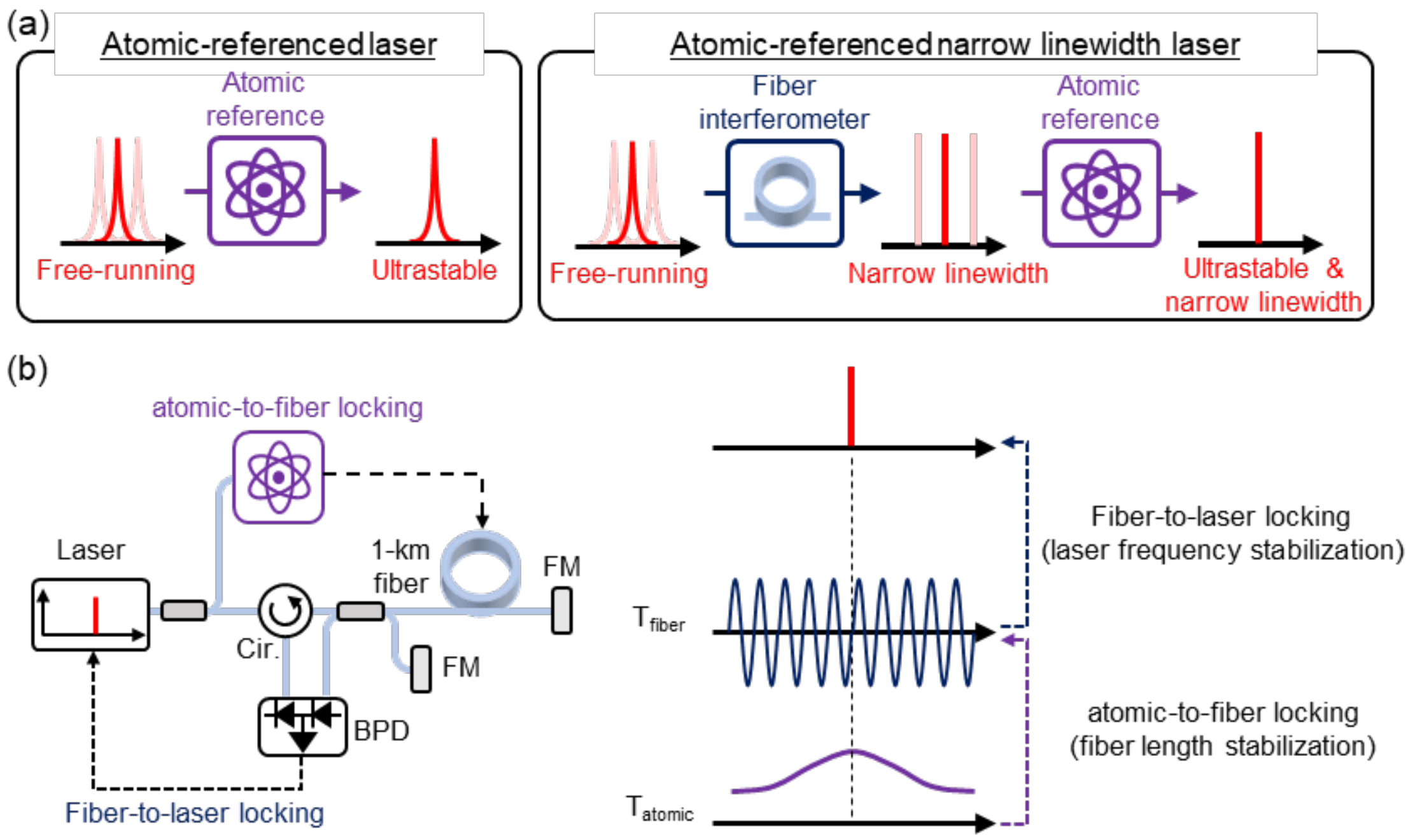

**Figure 1. Concept and operating principle of the atomic-referenced narrow-linewidth laser. a,** Comparison between a conventional atomic-referenced laser and the proposed scheme. In the conventional approach, the laser is directly locked to an atomic transition, providing long-term frequency anchoring but limited linewidth reduction. In contrast, the proposed scheme combines a fiber interferometer for ultra-narrow linewidth generation with an atomic transition that defines the absolute frequency reference. **b,** Operating principle of the hierarchical stabilization scheme. The laser is first stabilized to a fiber-delay interferometer for short-term spectral purification, while the interferometer is subsequently referenced to an atomic transition to suppress long-term drift. Cir, circulator; FM, Faraday mirror; BPD, balanced photodetector; $T_{fiber}$, interferometer transmission; $T_{atomic}$, transmission of the atomic reference signal.

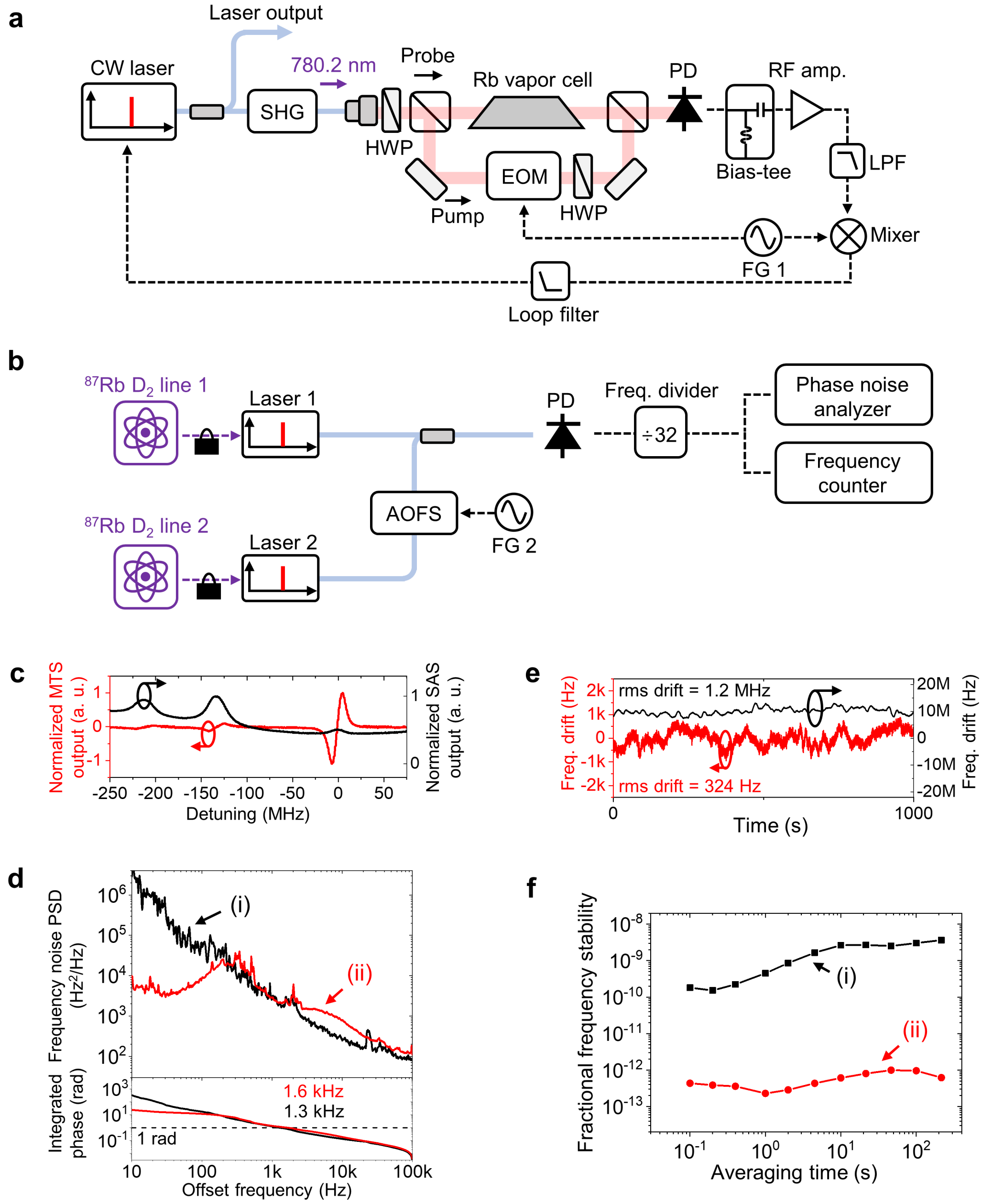

**Figure 2. Atomic-referenced laser and its performance. a,** Schematic of the $^{87}$Rb $D_2$ line-based MTS system used for atomic referencing. **b**, Schematic of the optical beat-note measurement for performance evaluation. **c,** Comparison of SAS and MTS signals as a function of detuning. **d,** Frequency noise PSDs of the free-running laser (curve (i)) and the atomic-referenced laser (curve (ii)). **e,** Frequency drift of the free-running laser (curve (i)) and the atomic-referenced laser (curve (ii)). **f,** Fractional frequency stability of the free-running laser (curve (i)) and the atomic-referenced laser (curve (ii)). SHG, second harmonic generation module; EOM, electro-optic modulator; PD, Photodetector; FG, function generator.

**a**

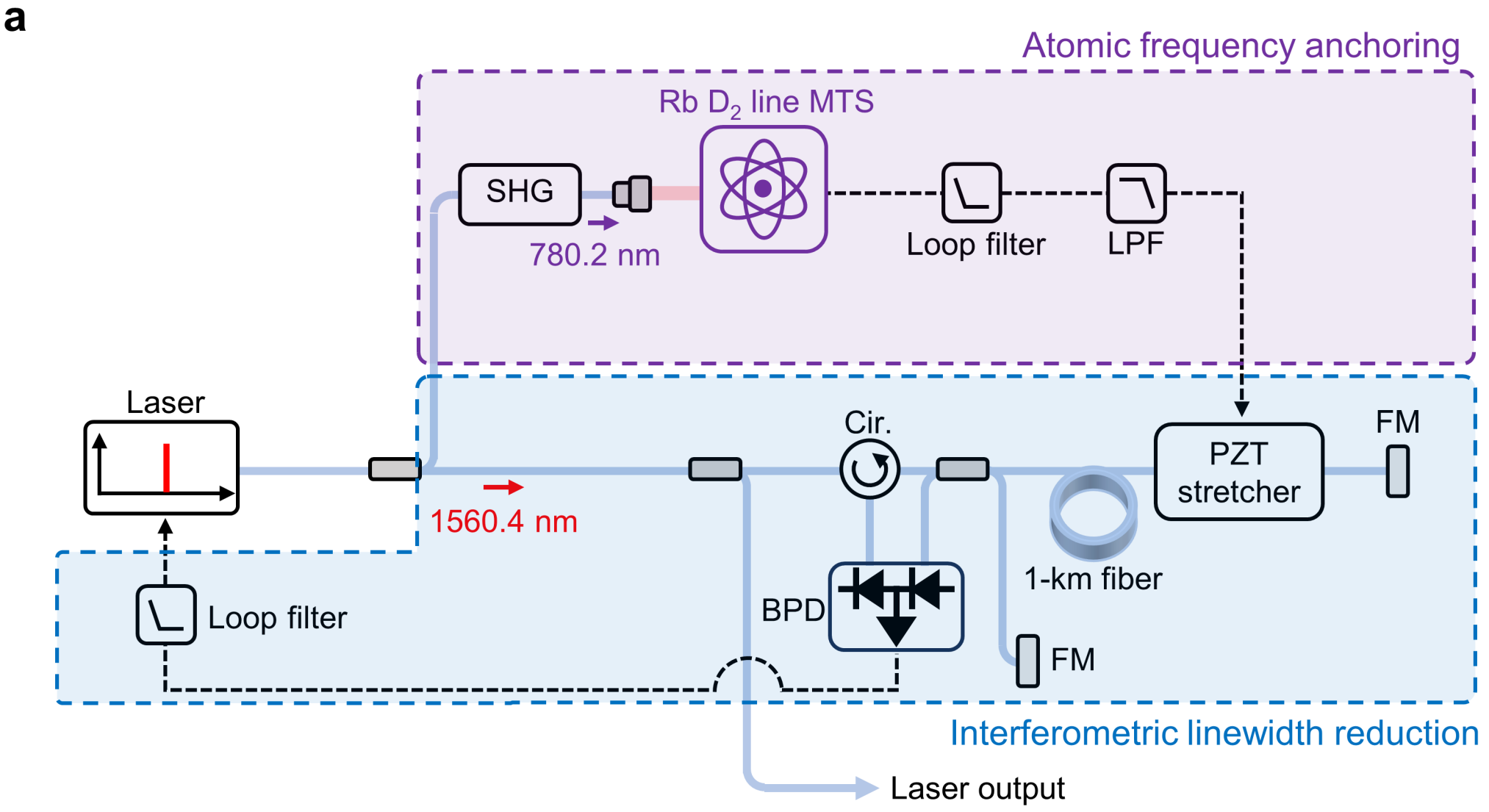

**b**

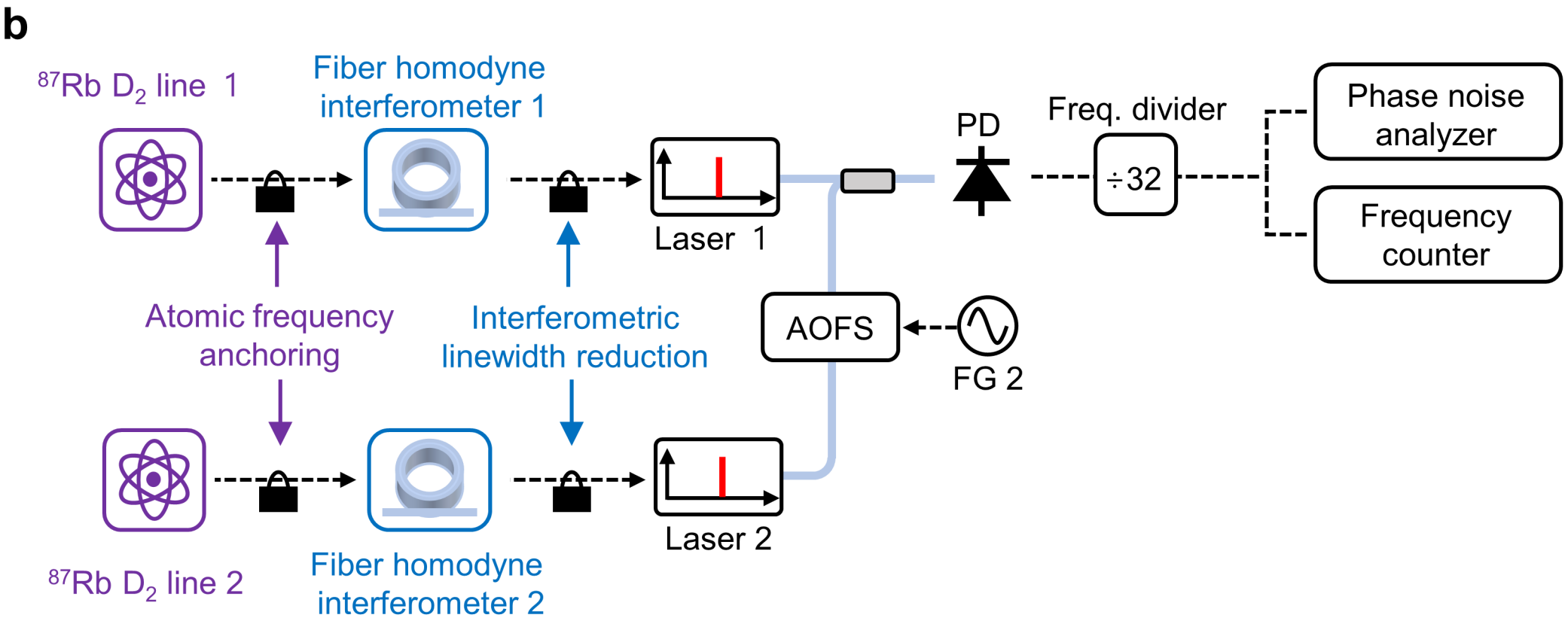

**Figure 3. Experimental implementation of the atomic-referenced narrow-linewidth laser. a,** Schematic of the hierarchical stabilization system combining interferometric linewidth reduction and atomic frequency anchoring. **b**, Schematic of the optical beat-note measurement for performance evaluation. Cir, circulator; FM, Faraday mirror; BPD, balanced photodetector; MTS, modulation transfer spectroscopy; LPF, low-pass filter; AOFS, acousto-optic frequency shifter; FG, function generator; PD, photodiode.

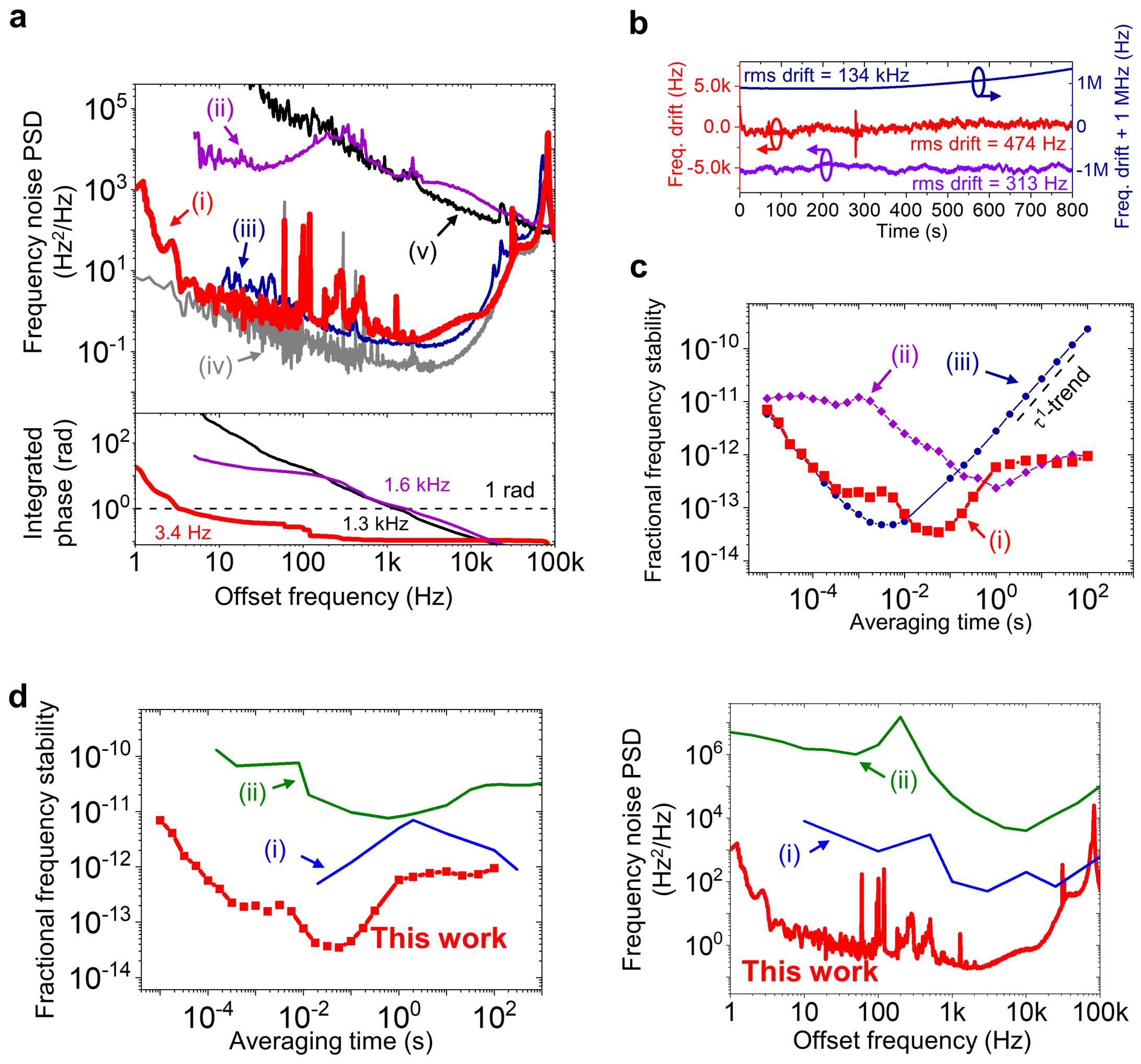

**Figure 4. Short- and long-term performance of the atomic-referenced narrow-linewidth laser. a,** Frequency noise PSDs of the proposed atomic-referenced narrow-linewidth laser (curve (i)), atomic-referenced laser (curve (ii)), fiber-stabilized laser (curve (iii)), in-loop interferometric signal (curve (iv)), and free-running laser (curve (v)). The corresponding integrated phase indicates a linewidth of 3.4 Hz, compared to ~1.6 kHz and ~1.3 kHz for the atomic-referenced and free-running lasers, respectively. **b,** Frequency drift of the atomic-referenced narrow-linewidth laser (red), atomic-referenced laser (purple), and fiber-stabilized laser (navy). For clarity, the purple and navy traces are vertically offset by -5 kHz and +1 MHz, respectively. **c,** Fractional frequency stability of the atomic-referenced narrow-linewidth laser (curve (i), red), atomic-referenced laser (curve (ii), purple), and fiber-stabilized laser (curve (iii), navy). At short averaging times, the stability follows that of the fiber-stabilized laser, while at longer times it converges to that of the atomic-referenced laser, indicating successful combination of short-term spectral purity and long-term frequency stability. The black dashed line indicates a $\tau^1$ dependence. **d,** Performance comparison of fractional frequency stability (left) and frequency noise PSD (right) with previously reported systems in refs. 37 (curve (i)) and 38 (curve (ii)).